\documentclass[reprint,showpacs,usenames,dvipsnames,aps,prl]{revtex4-1}
\usepackage{graphicx,color,epsfig,rotate}
\usepackage{times,xspace}
\usepackage{amsbsy,amssymb,amsmath,bm}
\usepackage{dcolumn}
\usepackage[normalem]{ulem}

\begin{document}

\title{Successive magnetic field-induced transitions and colossal magnetoelectric effect in Ni$_{3}$TeO$_{6}$}

\author{Jae~Wook~Kim$^{1,2}$}
\thanks{These authors contributed equally to the present work.}
\author{S.~Artyukhin$^{3}$}
\thanks{These authors contributed equally to the present work.}
\author{E.~D.~Mun$^{1}$}
\author{M.~Jaime$^{1}$}
\author{N.~Harrison$^{1}$}
\author{A.~Hansen$^{1}$}
\author{J.~J.~Yang$^{2}$}
\author{Y.~S.~Oh$^{2}$}
\author{D.~Vanderbilt$^{3}$}
\author{V.~S.~Zapf$^{1}$}
\author{S.-W.~Cheong$^{2}$}

\affiliation{$^1$Los Alamos National Laboratory, Los Alamos, NM 87545, USA}
\affiliation{$^2$Rutgers Center for Emergent Materials and Department of Physics and Astronomy, Rutgers University, Piscataway, NJ 08854, USA}
\affiliation{$^3$IAMDN and Department of Physics and Astronomy, Rutgers
University, Piscataway, NJ 08854, USA}

\date{\today}

\begin{abstract}
%
We report the discovery of a metamagnetic phase transition in a polar antiferromagnet Ni$_3$TeO$_6$ that occurs at 52~T.
The new phase transition accompanies a colossal magnetoelectric effect, with a magnetic-field-induced polarization change of 0.3~$\mu$C/cm$^2$, a value that is 4~times larger than for the spin-flop transition at 9~T in the same material, and also comparable to the largest magnetically-induced polarization changes observed to date.
Via density-functional calculations we construct a full microscopic model that describes the data.
We model the spin structures in all fields and clarify the physics behind the 52~T transition.
The high-field transition involves a competition between multiple different exchange interactions which drives the polarization change through the exchange-striction mechanism.
The resultant spin structure is rather counter-intuitive and complex, thus providing new insights on design principles for materials with strong magnetoelectric coupling.

\end{abstract}

\maketitle

Magnetoelectric (ME) multiferroics have been extensively studied recently to understand the mechanisms responsible for cross-coupling between magnetism and ferroelectricity, which is at the heart of their promise for application in multifunctional devices \cite{Spaldin2005,Eerenstein2006,Cheong2007,Kimura2007,Khomskii2009,Arima2011}.
In this class of materials, at least three mechanisms are known to induce ferroelectric polarization ($P$) upon magnetic order: (1) the spin current or inverse Dzyaloshinskii-Moriya (DM) interaction in a spin-cycloidal structure which is mediated by anti-symmetric exchange \cite{Katsura2005,Mostovoy2006,Sergienko2006}, (2) the symmetric exchange-striction mechanism between parallel or anti-parallel alignment of spins \cite{Choi2008}, and (3) the hybridization between metal $d$- and ligand $p$-orbitals that is modulated by spin direction \cite{Arima2007}.
A majority of ME couplings that have been studied to date involve mechanisms (1) and (3).
However, the symmetric exchange mechanism (2) can, in principle, also lead to large ME effects.

Another route to a large ME effect is to consider magnetic systems that have a polar structure.
This condition meets the prerequisites for the ME effect, i.e., coexistence of broken spatial-inversion-symmetry and time-reversal symmetry.
Often, these systems exhibit a non-polar to polar structural transition at high temperatures and magnetic ordering at lower temperatures.
A well-known example is BiFeO$_3$, with ferroelectric and antiferromagnetic transition temperatures at 1100~K and 653~K respectively \cite{Catalan2009}.
However, its ME effect is small compared to those of spin-driven ME materials.



\begin{figure}
\centering
\includegraphics[width=0.42\textwidth]{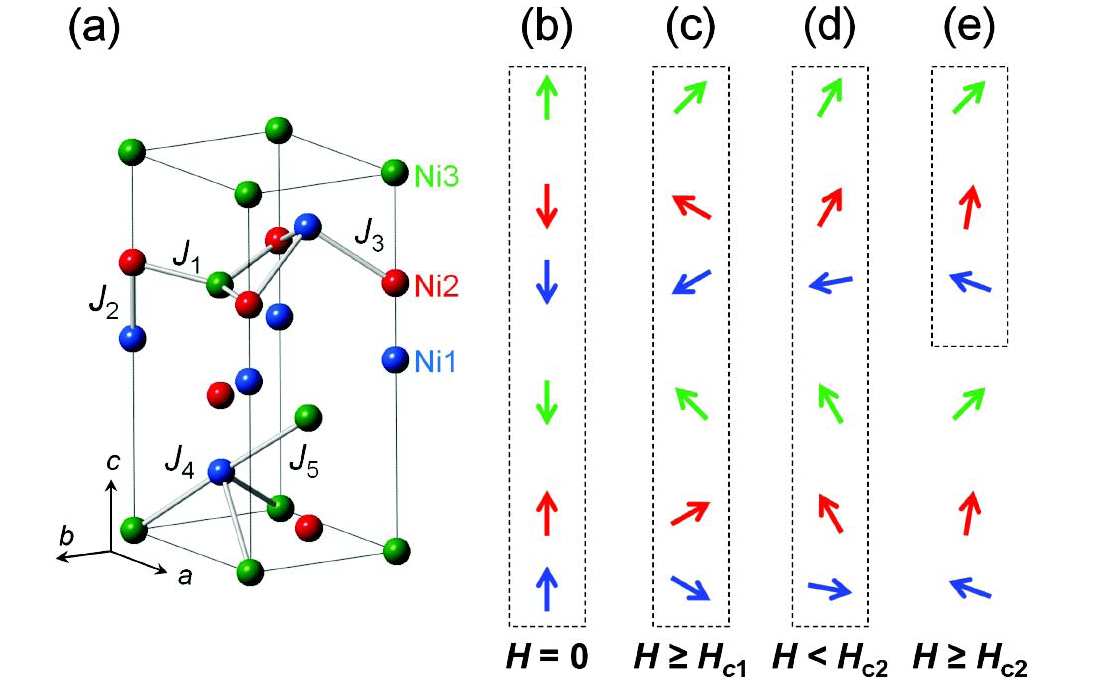}
\caption{
(a) Crystal structure of Ni$_3$TeO$_6$.
Only Ni ions are shown.
(b)-(e) Schematic spin structure along a $c$-axis chain at different magnetic fields applied along the $c$-axis.
The spins in the buckled honeycomb planes are aligned ferromagnetically in zero magnetic field \cite{Zivkovic2010}.
$J_i$ stands for the five nearest-neighbor exchange constants considered in the text.
Dotted boxes indicate the magnetic unit cell along the $c$-axis, which is doubled for (b)-(d) but not for (e).}
\label{structure}
\end{figure}

Recently, a ME effect has been observed in the corundum-related compound Ni$_3$TeO$_6$ (NTO).
It crystallizes in a polar $R3$ space-group with three Ni$^{2+}$ ions (3$d^8$, $S$~=~1) and a non-magnetic Te ion arranged along the $c$-axis in a unit cell (Fig.~\ref{structure}(a)) to form a linear chain with broken spatial-inversion symmetry.
This material is not ferroelectric but pyroelectric, i.e., the electric polarization cannot be switched by an external electric field.
However, in addition to its non-switchable electric polarization component due to the polar structure, it also shows a large magnetically-induced polarization.
In zero magnetic field ($H$), below the Ne\'el temperature $T_N$~=~52~K, NTO has a collinear antiferromagnetic (AFM) order,  $\uparrow\uparrow\uparrow\downarrow\downarrow\downarrow$ in the rhombohedral unit cell with spins aligned along the $c$-axis \cite{Zivkovic2010}.
The spatial inversion symmetry is broken and the electric polarization is pointing along the $c$-axis, as in Ca$_3$(Co,Mn)O$_6$ \cite{Choi2008,Kim2014}.
It was found that NTO undergoes a second-order spin-flop (SF) transition at a critical field $H_{c1}$~$\sim$~9~T, which accompanies a large ME effect \cite{Oh2014}.
Here, symmetric exchange-striction at the SF transition distorts the polar crystal structure to modify the electric polarization.
The ME coefficient ($\alpha$~$\equiv$~$\frac{dP}{dH}$) is as high as 1300~ps/m at $H_{c1}$, without any magnetic hysteresis.
However, a phenomenological model with only two magnetic sublattices was implemented to describe the polarization change at the SF transition \cite{Oh2014}, whereas a full description of NTO requires a model with three different Ni spins and all possible exchange interactions between them \cite{Zivkovic2010}.

In this Letter, we present a new phase transition in NTO discovered by high magnetic field study up to 92~T. 
The high field transition is accompanied by a colossal ME effect that is comparable to largest field-induced polarization changes reported to date \cite{Lee2013,Aoyama2014}.
We corroborate our experimental results with a microscopic model, using model parameters extracted from extensive density-functional theory (DFT) calculations, thereby predicting the magnetic structure at all magnetic fields.
Multiple exchange interactions are found to contribute to the field-dependence of electric polarization via an exchange-striction mechanism.

\begin{figure}
\centering
\includegraphics[width=0.5\textwidth]{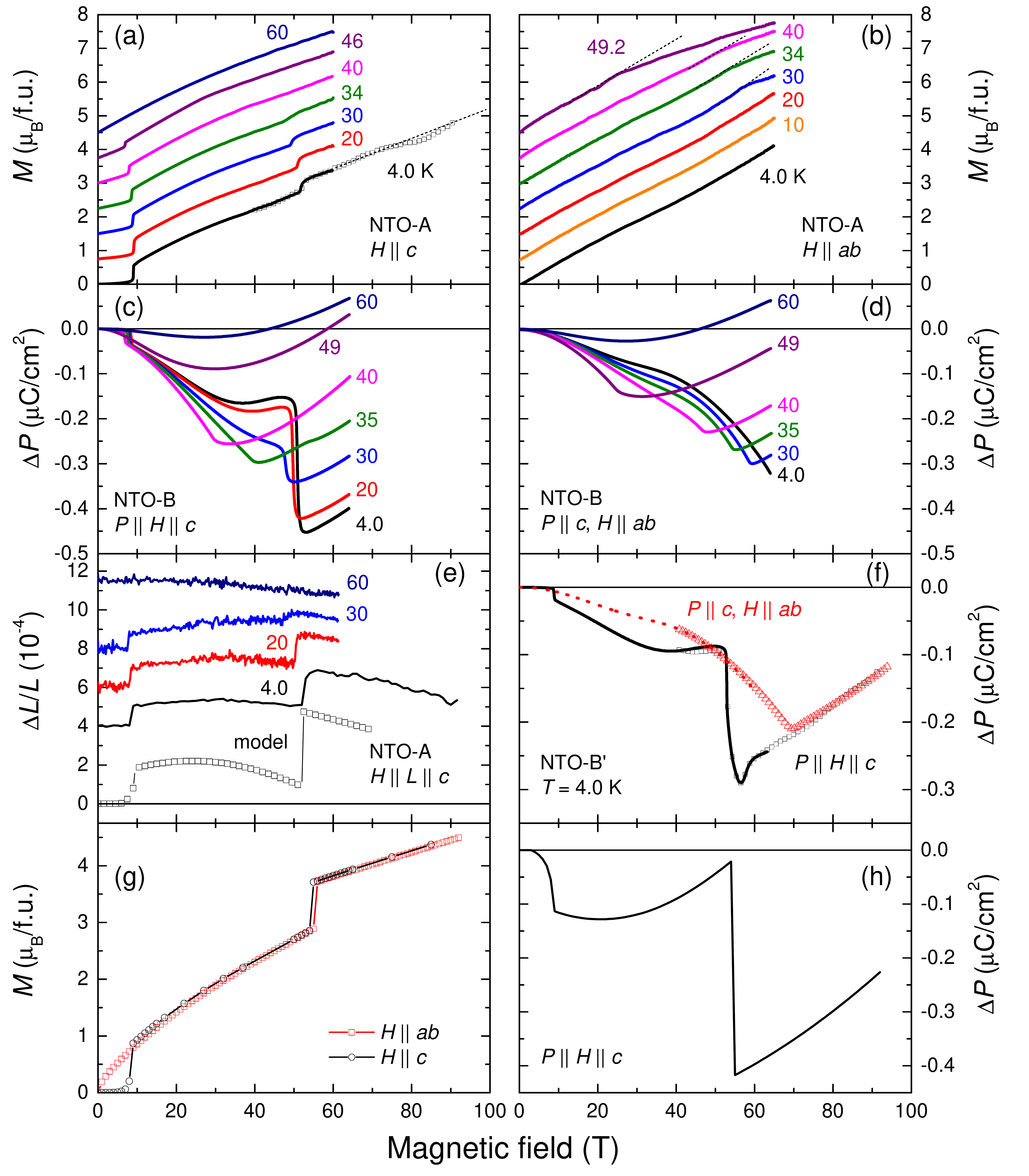}
\caption{
(a), (b) Magnetization and (c), (d) change of electric polarization ($\Delta$$P$) of Ni$_3$TeO$_6$ for magnetic fields applied along different crystalline axes as indicated.
(e) $c$-axis magnetostriction as a function of magnetic field applied along the $c$-axis.
Solid lines are experimental data taken under isothermal condition and open squares are from model calculations based on parameters shown in Table~\ref{table}.
(f) $\Delta$$P$ measured up to 92~T for $H$~$\parallel$~$ab$ and $H$~$\parallel$~$c$.
(g), (h) The magnetic field-dependence of $M$ and $\Delta$$P$ along the $c$-axis in $H$~$\parallel$~$c$ configuration obtained from the model calculations described in text.
Dashed lines in (a), (b) are guides for the eyes.
(a), (b) magnetization and (e) magnetostriction curves are shifted for clarity.
(e) Jump in magnetostriction at $H_{c2}$ shows a slight difference in magnitude between different types of magnet used, possibly due to the variation in magnetic field sweep rate.
(f) Lines and open symbols denote data taken by using a capacitor-bank-driven and a hybrid pulse-magnet, respectively.
A single domain sample was used for (c), (d) while a multi-domain (as-grown) sample was used for (f).
}
\label{Hdep}
\end{figure}

Fig.~\ref{Hdep}(a) shows the magnetic field-dependence of the magnetization $M$ along the $c$-axis of NTO up to 92~T.
At 4~K a sharp SF transition is evident at $H_{c1}$~$\sim$~9~T, then $M$ increases quasi-linearly up to 20~T.
The linear extrapolation of the $M$($H$) data between 9 and 20~T has a zero intercept at $H$~=~0, consistent with a SF transition.
When $H$ is further increased above 30~T, the slope of the $M$($H$) curve decreases slightly.
At $H_{c2}$~=~52~T, a small jump in $M$ is observed.
With further increasing $H$, $M$ increases linearly up to 92~T.
The value of $M$ at 92~T is 4.8~$\mu_{B}$ per formula unit (f.u.), which is still smaller than the expected saturation magnetization $M_S$~=~6~$\mu_{B}$/f.u. for three Ni$^{2+}$ ions with $S$~=~1 (assuming gyromagnetic ratio of 2).
By assuming that there are no other magnetic field-induced transitions, linear extrapolation of the $M$($H$) curve to the expected $M_S$ value gives a saturation magnetic field $H_{S}$ of 120$\pm$10~T.
The in-plane $M$($H$) data (Fig.~\ref{Hdep}(b)) below 30~K show a monotonic increase under magnetic field.
At 30~K, the $M$($H$) curve shows a cusp at $H$~=~60~T which decreases upon further warming.
The most striking feature of the high-field transition in NTO is the colossal change of $P$ at $H_{c2}$ and reversal of the $\Delta$$P$($H$) slope at higher fields.
Fig.~\ref{Hdep}(c) shows the change of $c$-axis electric polarization $\Delta$$P$~$\equiv$~$P$($H$)~--~$P$($H$=0) as a function of magnetic field applied along the $c$-axis.
In this configuration, $\Delta$$P$($H$) curve shows a step at $H_{c1}$ as previously reported \cite{Oh2014}.
When the magnetic field is further increased at 4~K, $P$ slightly increases, and then shows a sudden drop at 50~T, close to $H_{c2}$, with a minimum at 53~T.
The overall $\Delta$$P$ at $H_{c2}$ reaches 0.3~$\mu$C/cm$^2$ at 4~K.
Counter-intuitively, $\Delta$$P$ at $H_{c2}$ is $\sim$~10 times larger than that at $H_{c1}$, whereas the change of $M$ at $H_{c2}$ is almost two times smaller than that at $H_{c1}$.
When the magnetic field is further increased above 53~T, $P$ increases linearly up to 65~T.
In contrast, the $c$-axis polarization measured under in-plane magnetic field does not show any sharp jump, but only a smooth reversal in slope which shifts to lower magnetic field as the temperature is increased (Fig.~\ref{Hdep}(d)), concurrent with a cusp in the in-plane $M$($H$) curve (Fig.~\ref{Hdep}(b)).
The lattice also responds sensitively to the magnetic field at these transitions, as shown in the $c$-axis magnetostriciton $\Delta$$L$/$L$ measurements (Fig.~\ref{Hdep}(e)).

We further explored $\Delta$$P$($H$) up to 92~T, with a different, multi-domain sample (Fig.~\ref{Hdep}(f)).
When the magnetic field is applied along the $c$-axis, $P$ increase linearly above $H_{c2}$, up to 92~T.
A linear extrapolation of $\Delta$$P$($H$) curve above $H_{c2}$ gives $\Delta$$P$($H$)~=~0 at 120$\pm$5~T, 
consistent with the expected saturation magnetic field from the $M$($H$) curve.
This implies that the magnetically-induced electric polarization is no longer active when the system is in the fully saturated phase.
At elevated temperatures, the sharp changes of $P$ at $H_{c2}$ are still observed, although its magnitude and transition field decreases up to 30~K, above which the sharp drop is replaced with a slope change in the $\Delta$$P$($H$) curve.
When the magnetic field is applied along the $ab$-plane, the $\Delta$$P$($H$) curve shows only a sharp reversal of slope at 70~T at 4~K.
Above 70~T, two $\Delta$$P$($H$) curves measured in different configurations coincide with each other, suggesting an isotropic magnetic behavior above this field.

The field-induced $\Delta$$P$ value of 0.3~$\mu$C/cm$^2$ (Fig.~\ref{Hdep}(c)) and the high ME coefficient value \cite{Supplement} at $H_{c2}$ in NTO are among the largest observed in magnetoelectric systems \cite{Lee2013,Aoyama2014,Chun2010}.
All spin-driven ME materials exhibit a change of $P$ when a field-induced phase transition or spin-reorientation occurs \cite{Cheong2007,Kimura2007,Arima2011}.
However, for most of them, $\Delta$$P$($H$) is typically less than 0.01~$\mu$C/cm$^2$.
Thus, NTO is a prototypical example where the polar symmetry and the additional polarization coupled to the magnetically ordered state give rise to a large ME coupling.

The field-dependence of $\Delta$$P$ is similar to the case of BiFeO$_3$ along certain directions in that the $\Delta$$P$($H$) curve shows a step-like feature and a slope change \cite{Kadomtseva2004,Tokunaga2010}.
However, in the case of BiFeO$_3$ the field-dependent behavior is due to a phase transition from a spin-cycloid to a canted-AFM state, whereas we do not find any evidence of spin-cycloid or spiral structure in NTO.
In addition, the spin-spiral state in NTO is only allowed in the $ab$-plane by symmetry, where it cannot contribute to $c$-axis polarization.



We note that $\Delta$$P$ displays no electric field-dependence at either magnetic transitions \cite{Supplement}, suggesting that NTO is not a type-II multiferroic where magnetic order induces ferroelectricity \cite{Khomskii2009}.
Rather, the magnetic order modifies the existing electric polarization associated with the polar space-group which is established already at very high temperature (a possible ferroelectric $T_C$ of $\sim$~1000~K was reported \cite{Ivanov2013b}).

\begin{figure}
\centering
\includegraphics[width=0.45\textwidth]{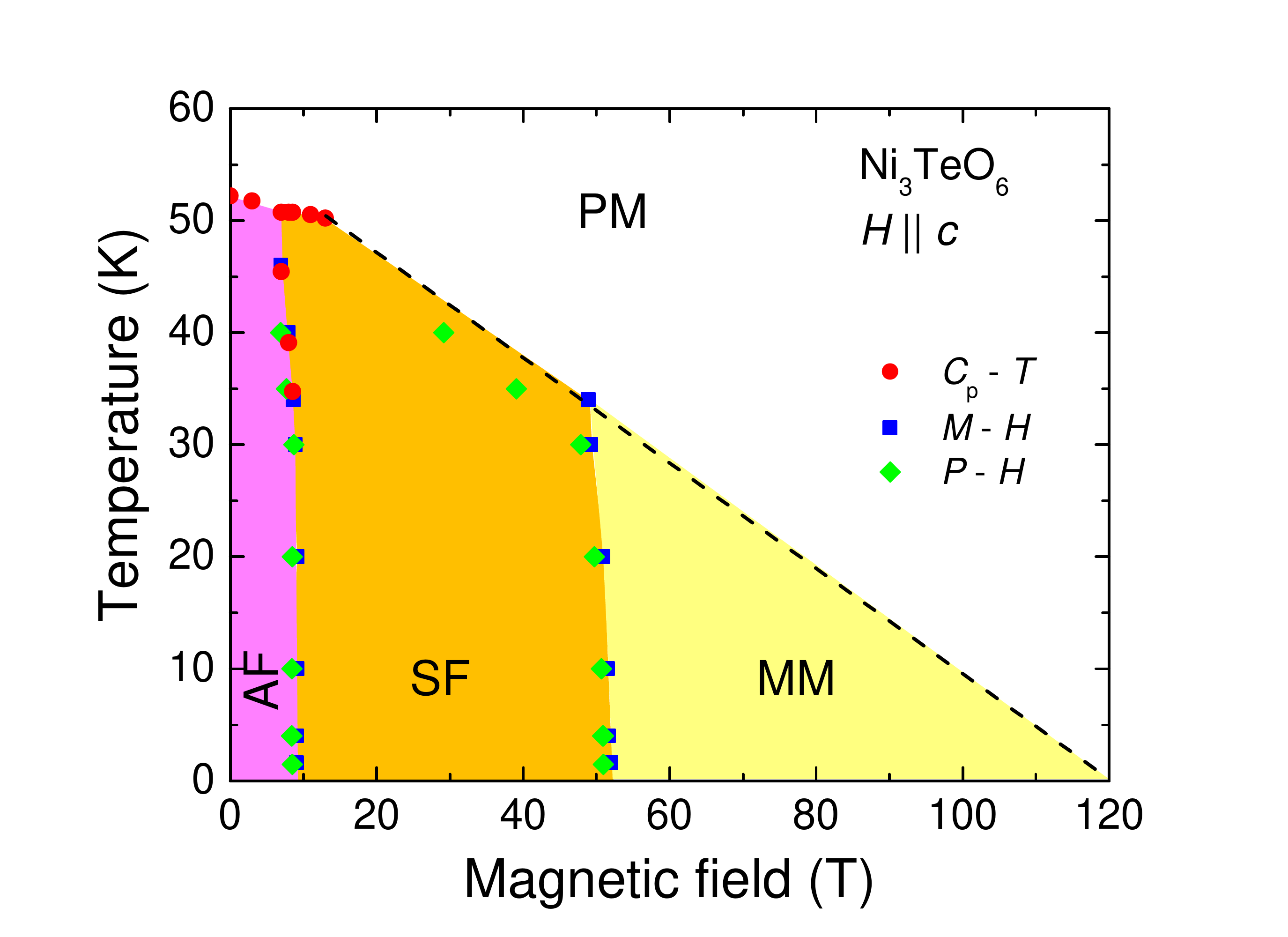}
\caption{
$H$-$T$ phase diagram of Ni$_3$TeO$_6$ with magnetic field applied along the $c$-axis determined by peak positions in d$M$/d$H$, $\alpha$~$\equiv$~d$P$/d$H$, and specific heat (see Supplement \cite{Supplement}) data.
Dashed line is a guide for the eyes.}
\label{HT}
\end{figure}

Using our experimental results, we construct an $H$-$T$ phase diagram for NTO in magnetic field along the $c$-axis, shown in Fig.~\ref{HT}.
We observe three ordered phases below $T_N$: AFM, SF, and metamagnetic phase (MM) which are separated by phase boundaries, determined by $M$($H$) and $\Delta$$P$($H$) curves, that are almost vertical at low temperatures.
The high-temperature phase boundary between the paramagnetic (PM) and the ordered phase (dotted line in Fig.~\ref{HT}) extrapolates linearly to 120~T at $T$~=~0, again, consistent with the extrapolation of the $M$($H$) and $\Delta$$P$($H$) curves to their expected saturation and zero values, respectively.

Turning now to the theoretical modeling of this material, we note that the phenomenological description of magnetism in NTO \cite{Oh2014} is applicable in the vicinity of the low-field SF transition, but may not be accurate away from it.
In order to study magnetic transitions in the whole magnetic field range, we use a simplified microscopic model with the Hamiltonian
\begin{equation}
\label{eq1}
{\cal H} = \sum_{i,j} J^{(ij)} \vec{S}_i \cdot \vec{S}_j
+\sum_i \left[ - K_{2,i} (S_{i,z})^2 - \vec{H} \cdot \vec{S}_i \right],
\end{equation}
where we model the Ni spins $\vec S_i$ ($S$~=~1) classically, and $i,j$ 
run over all Ni sites in the lattice.
The exchange constants between Ni spins are $J^{(ij)}$, taking particular values $J_1\dots J_5$ for the bonds $(ij)$ indicated in Fig.~\ref{structure}.
We neglected anisotropic exchanges as well as next-nearest-neighbor Heisenberg exchanges, since the second transition appears in the model without them.
The terms with $K_2$ and $\vec{H}$ model single-ion easy-$c$-axis anisotropy and the coupling to a uniform magnetic field, respectively.

The exchange constants depend on the ionic coordinates, and as a result, the ions shift in response to spin reorientations in such a way as to strengthen the exchanges that favor the existing spin arrangement.
These shifts of charged ions in a polar structure result in a change of $P$, which, assuming Heisenberg exchange-striction dominates, can be expressed (neglecting higher-order terms containing $(\vec S_i \cdot \vec S_j)^2$) as
\begin{equation}
\label{eq2}
\Delta P_c  = \sum_n \alpha_n \, \vec{S}_n \cdot \vec{S}_n' ,
\end{equation}
where $\vec S_n$ and $\vec S_n'$ are the spins connected by the exchange interaction $J_n$, and the $\alpha_n$ are exchange-striction parameters.
We use Eq.~(\ref{eq2}) to model the dependence of the polarization $P$ on the spin configuration. 

The coefficients $\alpha_n$~=~$\alpha_\textrm{n,ion}$~+~$\alpha_\textrm{n,el}$, with the two terms describing the polarization contributions due to ion shifts and deformations of electronic orbitals, respectively, are calculated using DFT \cite{Supplement}.
Similarly, the magnetostriction -- the change of the sample size under an applied magnetic field -- is described by the equation, analogous to (\ref{eq2}): ${\Delta L_c/L_c  = \sum_n \lambda_n \, \vec{S}_n \cdot \vec{S}_n'}$.

The determination of exchange constants $J_n$ is a difficult problem.
The values calculated previously using DFT \cite{Wu2010} give a non-collinear ground state when the energy is minimized within a magnetic unit cell at realistic
values of $K_2$.
We have found that the exchange constants estimated using the PBE0 hybrid functional approximation to DFT \cite{PBE0} give the correct ground state and reproduce the second transition.
We have then adjusted these constants to better fit the experimental $M$($H$) data measured along the $c$-axis.
The resulting $J_n$ parameters are summarized in Table~\ref{table} along with the exchange-striction constants $\alpha_n$ and magnetostriction parameters $\lambda_n$, calculated using DFT~+~$U$, as described in the Supplement \cite{Supplement}.

With these model parameters in hand, we computed the changes in the spin arrangement under magnetic field applied along the $c$-axis; the results are shown in Fig.~\ref{structure}(b-e).
The resulting magnetization and polarization curves, presented in Fig.~\ref{Hdep}(g,h), are in qualitative agreement with experiments.
We checked that the calculated transition sequence and spin structures did not change significantly with tuning of the exchange constants, suggesting an absence of competing phases.
Our confidence in our correct identification of the phase transitions is further reinforced by the agreement of the calculated and measured magnetization and polarization curves.

In the textbook SF transition for a two-sublattice antiferromagnet, the exchange favors the collinear state \cite{LL8c50}.
Surprisingly, in NTO the situation is the opposite -- the AFM exchange $J_5$ actually favors the canted state.
Ferromagnetic exchange $J_2$ favors the collinear state, in which the spins of Ni1 and Ni2 are parallel to each other, as shown in Fig.~\ref{Hdep}.
The evolution of energy contributions from different exchanges under the applied magnetic field is illustrated in the Supplement \cite{Supplement}.
In the canted state, above $H_{c1}$, the spins of Ni1 and Ni2 are no longer parallel thus losing energy on $J_2$, but this canting allows a gain in energy from other exchange interactions, while gaining Zeeman energy from the canting of the Ni2 and Ni3 spins along the magnetic field, as shown in Fig.~\ref{Hdep}(c).

As the magnetic field is increased further above $H_{c1}$, the spin of the Ni2 rotates.
At high-fields the $c$-component of the spins is pinned by the field, while the $ab$-plane components are chosen to minimize the energy (Eq.~(\ref{eq1})).
This is similar to the energy of the collinear state, except that the $ab$-component is not constrained to have the unit length.
That is why the state that results above $H_{c1}$, with six spins pointing, $\rightarrow$0$\leftarrow\rightarrow$0$\leftarrow$, differs from the zero-field state, $\uparrow\uparrow\downarrow\downarrow\downarrow\uparrow$.
Above $H_{c2}$, the spin of the Ni1 cants further along the magnetic field, and simultaneously the $ab$-plane components of the spins in half of the magnetic unit cell reverse in order to gain energy on the antiferromagnetic $J_5$ exchange, acting between Ni1 and Ni3 spins from the neighboring crystallographic unit cells.
At the same time the energy contributions from all the other exchanges increase, as evidenced by the total energy calculation \cite{Supplement}.
A reversal of the $ab$-plane components of the spins in every second crystallographic unit cell leads to a large change of the magnetically-induced electric polarization and strains, as seen in Fig.~\ref{Hdep}(c,e,h).
We can see a restoration of translational symmetry along the $c$-axis that was broken by AFM ordering.

\begin{table}
\caption{\label{table} Exchange ($J_n$), easy-axis anisotropy constants ($K_2$), exchange-striction parameters for electronic ($\alpha_{n,\text{el}}$) and ionic contributions ($\alpha_{n,\text{ion}}$) to the electric polarization, and magnetostriction parameters ($\lambda_n$) estimated using DFT calculations and adjusted to ensure an AFM ground state.
}
\begin{tabular}{|c | c c c c c c|}
\hline  $n$ & 1 & 2 & 3 & 4 & 5& $K_2$ (meV) \\
\hline
$J_{n}^\text{GGA}$(meV) & -0.6 & -3.1 & 2.2 & 4.2 & 1.0& \\
$J_{n}^\text{GGA-adj}$(meV) & -0.6 & -3.1 & 2.2 & 4.2 & 0.69& \\
$\alpha_{n,\text{el}}$ & 0.25 & 3.3 & -0.1 & -3.0 & -0.8& \\
$\alpha_{n,\text{ion}}$ & -2.4 & -1.6 & -2.0 & 11.2 & 6.2& \\
\hline
$\lambda_n\times 10^6$& 2.1& -1.6& -4.3& -8.9& -15.3&\\
\hline
$J_n^\text{PBE0}$(meV)&-1.13& -2.97& 0.79& 2.76& 0.32&0.05\\
$J_n^\text{PBE0-adj}$(meV)&-0.69&-3.63&0.76&3.26&0.65&0.1\\\hline
\end{tabular}
\end{table}

While the magnetic single-ion anisotropy plays an important role for the SF transition, the second transition at $H_{c2}$ is controlled by the magnetic exchanges and the external magnetic field, and is not sensitive to the single-ion anisotropy, thus it is not a classical SF transition.
Our model predicts the transition at $H_{c2}$ for both $H$~$\parallel$~$c$ and $H$~$\parallel$~$ab$.
Experimentally, however, no sharp transitions are observed in $H$~$\parallel$~$ab$, but there exists a cusp in the $P$($H$) at around 70~T at $T$~=~4~K, suggesting that the transition is of the second-order.
As seen in Fig.~\ref{Hdep}(a,b,f), the magnetization curves and $c$-axis electric polarization with magnetic field applied different directions nearly coincide above the transition, implying that the magnetic states for $H$~$\parallel$~$c$ and $H$~$\parallel$~$ab$ above the transition are similar.
This discrepancy between the theory and experiment is partly due to the presently neglected DM interactions and symmetric anisotropic exchanges.
In particular, our DFT calculations indicate the presence of strong Ising-type anisotropies, which will modify the phase diagram for the $c$-axis ordered structures much more than the $ab$-plane ones.
A detailed description of this aspect requires the introduction of additional parameters in our model, and additional experiments are being conducted in order to determine the parameters reliably.
This work will be reported elsewhere.

In summary, a high-field study of NTO reveals the presence of a second spin reorientation transition well above the SF transition at 9~T.
The high-field transition is first-order at 52~T for $H$~$\parallel$~$c$, and is second-order at 70~T for $H$~$\parallel$~$ab$ at base temperature.
Our theoretical analysis suggests that this high-field transition is governed by the competition between the Zeeman and exchange energies, and entails the reversal of the $ab$-plane component of half of the spins, leading via the Heisenberg exchange-striction to a change of electric polarization that is among the largest observed to date.
This spin reorientation results in restoration of translational symmetry along the $c$-axis that was broken by AFM ordering.
Our calculations also identify particular exchange interactions that are responsible for the stabilization of the magnetic phases in NTO.
Furthermore, compared to its isostructural compounds (Mn$_3$TeO$_6$ and Co$_3$TeO$_6$), which show multiferroic behavior via antisymmetric exchange \cite{Ivanov2011,Hudl2011,Li2012}, NTO exhibits an exchange-striction-driven polarization response that is almost two orders of magnitude larger, thereby demonstrating a unique behavior in this class of materials.
We propose that this scenario can form the basis of a new strategy for the design of materials with large ME effects.

\begin{acknowledgments}
The NHMFL Pulsed Field Facility is supported by the NSF, the U.S. D.O.E., and the State of Florida through NSF cooperative grant DMR-1157490.
Work at Los Alamos National Laboratory was supported by the U.S. D.O.E. BES project ``Science at 100 tesla" (BES FWP LANLF100).
The work at Rutgers was supported by the NSF grants DMREF-1104484, DMREF 12-33349, and the Rutgers IAMDN.
\end{acknowledgments}

\bibliography{NTO-jw}

\begin{thebibliography}{41}%
\makeatletter
\providecommand \@ifxundefined [1]{%
 \@ifx{#1\undefined}
}%
\providecommand \@ifnum [1]{%
 \ifnum #1\expandafter \@firstoftwo
 \else \expandafter \@secondoftwo
 \fi
}%
\providecommand \@ifx [1]{%
 \ifx #1\expandafter \@firstoftwo
 \else \expandafter \@secondoftwo
 \fi
}%
\providecommand \natexlab [1]{#1}%
\providecommand \enquote  [1]{``#1''}%
\providecommand \bibnamefont  [1]{#1}%
\providecommand \bibfnamefont [1]{#1}%
\providecommand \citenamefont [1]{#1}%
\providecommand \href@noop [0]{\@secondoftwo}%
\providecommand \href [0]{\begingroup \@sanitize@url \@href}%
\providecommand \@href[1]{\@@startlink{#1}\@@href}%
\providecommand \@@href[1]{\endgroup#1\@@endlink}%
\providecommand \@sanitize@url [0]{\catcode `\\12\catcode `\$12\catcode
  `\&12\catcode `\#12\catcode `\^12\catcode `\_12\catcode `\%12\relax}%
\providecommand \@@startlink[1]{}%
\providecommand \@@endlink[0]{}%
\providecommand \url  [0]{\begingroup\@sanitize@url \@url }%
\providecommand \@url [1]{\endgroup\@href {#1}{\urlprefix }}%
\providecommand \urlprefix  [0]{URL }%
\providecommand \Eprint [0]{\href }%
\providecommand \doibase [0]{http://dx.doi.org/}%
\providecommand \selectlanguage [0]{\@gobble}%
\providecommand \bibinfo  [0]{\@secondoftwo}%
\providecommand \bibfield  [0]{\@secondoftwo}%
\providecommand \translation [1]{[#1]}%
\providecommand \BibitemOpen [0]{}%
\providecommand \bibitemStop [0]{}%
\providecommand \bibitemNoStop [0]{.\EOS\space}%
\providecommand \EOS [0]{\spacefactor3000\relax}%
\providecommand \BibitemShut  [1]{\csname bibitem#1\endcsname}%
\let\auto@bib@innerbib\@empty
\bibitem [{\citenamefont {Spaldin}\ and\ \citenamefont
  {Fiebig}(2005)}]{Spaldin2005}%
  \BibitemOpen
  \bibfield  {author} {\bibinfo {author} {\bibfnamefont {N.~A.}\ \bibnamefont
  {Spaldin}}\ and\ \bibinfo {author} {\bibfnamefont {M.}~\bibnamefont
  {Fiebig}},\ }\href@noop {} {\bibfield  {journal} {\bibinfo  {journal}
  {Science}\ }\textbf {\bibinfo {volume} {309}},\ \bibinfo {pages} {391}
  (\bibinfo {year} {2005})}\BibitemShut {NoStop}%
\bibitem [{\citenamefont {Eerenstein}\ \emph {et~al.}(2006)\citenamefont
  {Eerenstein}, \citenamefont {Mathur},\ and\ \citenamefont
  {Scott}}]{Eerenstein2006}%
  \BibitemOpen
  \bibfield  {author} {\bibinfo {author} {\bibfnamefont {W.}~\bibnamefont
  {Eerenstein}}, \bibinfo {author} {\bibfnamefont {N.~D.}\ \bibnamefont
  {Mathur}}, \ and\ \bibinfo {author} {\bibfnamefont {J.~F.}\ \bibnamefont
  {Scott}},\ }\href@noop {} {\bibfield  {journal} {\bibinfo  {journal}
  {Nature}\ }\textbf {\bibinfo {volume} {442}},\ \bibinfo {pages} {759}
  (\bibinfo {year} {2006})}\BibitemShut {NoStop}%
\bibitem [{\citenamefont {Cheong}\ and\ \citenamefont
  {Mostovoy}(2007)}]{Cheong2007}%
  \BibitemOpen
  \bibfield  {author} {\bibinfo {author} {\bibfnamefont {S.-W.}\ \bibnamefont
  {Cheong}}\ and\ \bibinfo {author} {\bibfnamefont {M.}~\bibnamefont
  {Mostovoy}},\ }\href@noop {} {\bibfield  {journal} {\bibinfo  {journal} {Nat.
  Mater.}\ }\textbf {\bibinfo {volume} {6}},\ \bibinfo {pages} {13} (\bibinfo
  {year} {2007})}\BibitemShut {NoStop}%
\bibitem [{\citenamefont {Kimura}(2007)}]{Kimura2007}%
  \BibitemOpen
  \bibfield  {author} {\bibinfo {author} {\bibfnamefont {T.}~\bibnamefont
  {Kimura}},\ }\href@noop {} {\bibfield  {journal} {\bibinfo  {journal} {Annu.
  Rev. Mater. Res.}\ }\textbf {\bibinfo {volume} {37}},\ \bibinfo {pages} {387}
  (\bibinfo {year} {2007})}\BibitemShut {NoStop}%
\bibitem [{\citenamefont {Khomskii}(2009)}]{Khomskii2009}%
  \BibitemOpen
  \bibfield  {author} {\bibinfo {author} {\bibfnamefont {D.~I.}\ \bibnamefont
  {Khomskii}},\ }\href@noop {} {\bibfield  {journal} {\bibinfo  {journal}
  {Physics}\ }\textbf {\bibinfo {volume} {2}},\ \bibinfo {pages} {20} (\bibinfo
  {year} {2009})}\BibitemShut {NoStop}%
\bibitem [{\citenamefont {Arima}(2011)}]{Arima2011}%
  \BibitemOpen
  \bibfield  {author} {\bibinfo {author} {\bibfnamefont {T.}~\bibnamefont
  {Arima}},\ }\href@noop {} {\bibfield  {journal} {\bibinfo  {journal} {J.
  Phys. Soc. Jpn.}\ }\textbf {\bibinfo {volume} {80}},\ \bibinfo {pages}
  {052001} (\bibinfo {year} {2011})}\BibitemShut {NoStop}%
\bibitem [{\citenamefont {Katsura}\ \emph {et~al.}(2005)\citenamefont
  {Katsura}, \citenamefont {Nagaosa},\ and\ \citenamefont
  {Balatsky}}]{Katsura2005}%
  \BibitemOpen
  \bibfield  {author} {\bibinfo {author} {\bibfnamefont {H.}~\bibnamefont
  {Katsura}}, \bibinfo {author} {\bibfnamefont {N.}~\bibnamefont {Nagaosa}}, \
  and\ \bibinfo {author} {\bibfnamefont {A.~V.}\ \bibnamefont {Balatsky}},\
  }\href@noop {} {\bibfield  {journal} {\bibinfo  {journal} {Phys. Rev. Lett.}\
  }\textbf {\bibinfo {volume} {95}},\ \bibinfo {pages} {057205} (\bibinfo
  {year} {2005})}\BibitemShut {NoStop}%
\bibitem [{\citenamefont {Mostovoy}(2006)}]{Mostovoy2006}%
  \BibitemOpen
  \bibfield  {author} {\bibinfo {author} {\bibfnamefont {M.}~\bibnamefont
  {Mostovoy}},\ }\href@noop {} {\bibfield  {journal} {\bibinfo  {journal}
  {Phys. Rev. Lett.}\ }\textbf {\bibinfo {volume} {96}},\ \bibinfo {pages}
  {067601} (\bibinfo {year} {2006})}\BibitemShut {NoStop}%
\bibitem [{\citenamefont {Sergienko}\ and\ \citenamefont
  {Dagotto}(2006)}]{Sergienko2006}%
  \BibitemOpen
  \bibfield  {author} {\bibinfo {author} {\bibfnamefont {I.~A.}\ \bibnamefont
  {Sergienko}}\ and\ \bibinfo {author} {\bibfnamefont {E.}~\bibnamefont
  {Dagotto}},\ }\href@noop {} {\bibfield  {journal} {\bibinfo  {journal} {Phys.
  Rev. B}\ }\textbf {\bibinfo {volume} {73}},\ \bibinfo {pages} {094434}
  (\bibinfo {year} {2006})}\BibitemShut {NoStop}%
\bibitem [{\citenamefont {Choi}\ \emph {et~al.}(2008)\citenamefont {Choi},
  \citenamefont {Yi}, \citenamefont {Lee}, \citenamefont {Huang}, \citenamefont
  {Kiryukhin},\ and\ \citenamefont {Cheong}}]{Choi2008}%
  \BibitemOpen
  \bibfield  {author} {\bibinfo {author} {\bibfnamefont {Y.~J.}\ \bibnamefont
  {Choi}}, \bibinfo {author} {\bibfnamefont {H.~T.}\ \bibnamefont {Yi}},
  \bibinfo {author} {\bibfnamefont {S.}~\bibnamefont {Lee}}, \bibinfo {author}
  {\bibfnamefont {Q.}~\bibnamefont {Huang}}, \bibinfo {author} {\bibfnamefont
  {V.}~\bibnamefont {Kiryukhin}}, \ and\ \bibinfo {author} {\bibfnamefont
  {S.-W.}\ \bibnamefont {Cheong}},\ }\href@noop {} {\bibfield  {journal}
  {\bibinfo  {journal} {Phys. Rev. Lett.}\ }\textbf {\bibinfo {volume} {100}},\
  \bibinfo {pages} {047601} (\bibinfo {year} {2008})}\BibitemShut {NoStop}%
\bibitem [{\citenamefont {Arima}(2007)}]{Arima2007}%
  \BibitemOpen
  \bibfield  {author} {\bibinfo {author} {\bibfnamefont {T.}~\bibnamefont
  {Arima}},\ }\href@noop {} {\bibfield  {journal} {\bibinfo  {journal} {J.
  Phys. Soc. Jpn.}\ }\textbf {\bibinfo {volume} {76}},\ \bibinfo {pages}
  {073702} (\bibinfo {year} {2007})}\BibitemShut {NoStop}%
\bibitem [{\citenamefont {Catalan}\ and\ \citenamefont
  {Scott}(2009)}]{Catalan2009}%
  \BibitemOpen
  \bibfield  {author} {\bibinfo {author} {\bibfnamefont {G.}~\bibnamefont
  {Catalan}}\ and\ \bibinfo {author} {\bibfnamefont {J.}~\bibnamefont
  {Scott}},\ }\href@noop {} {\bibfield  {journal} {\bibinfo  {journal} {Adv.
  Mater.}\ }\textbf {\bibinfo {volume} {21}},\ \bibinfo {pages} {2463}
  (\bibinfo {year} {2009})}\BibitemShut {NoStop}%
\bibitem [{\citenamefont {\v{Z}ivkovi\'{c}}\ \emph {et~al.}(2010)\citenamefont
  {\v{Z}ivkovi\'{c}}, \citenamefont {Pr\v{s}a}, \citenamefont {Zaharko},\ and\
  \citenamefont {Berger}}]{Zivkovic2010}%
  \BibitemOpen
  \bibfield  {author} {\bibinfo {author} {\bibfnamefont {I.}~\bibnamefont
  {\v{Z}ivkovi\'{c}}}, \bibinfo {author} {\bibfnamefont {K.}~\bibnamefont
  {Pr\v{s}a}}, \bibinfo {author} {\bibfnamefont {O.}~\bibnamefont {Zaharko}}, \
  and\ \bibinfo {author} {\bibfnamefont {H.}~\bibnamefont {Berger}},\
  }\href@noop {} {\bibfield  {journal} {\bibinfo  {journal} {J. Phys.: Condens.
  Matter}\ }\textbf {\bibinfo {volume} {22}},\ \bibinfo {pages} {056002}
  (\bibinfo {year} {2010})}\BibitemShut {NoStop}%
\bibitem [{\citenamefont {Kim}\ \emph {et~al.}(2014)\citenamefont {Kim},
  \citenamefont {Kamiya}, \citenamefont {Mun}, \citenamefont {Jaime},
  \citenamefont {Harrison}, \citenamefont {Thompson}, \citenamefont
  {Kiryukhin}, \citenamefont {Yi}, \citenamefont {Oh}, \citenamefont {Cheong},
  \citenamefont {Batista},\ and\ \citenamefont {Zapf}}]{Kim2014}%
  \BibitemOpen
  \bibfield  {author} {\bibinfo {author} {\bibfnamefont {J.~W.}\ \bibnamefont
  {Kim}}, \bibinfo {author} {\bibfnamefont {Y.}~\bibnamefont {Kamiya}},
  \bibinfo {author} {\bibfnamefont {E.~D.}\ \bibnamefont {Mun}}, \bibinfo
  {author} {\bibfnamefont {M.}~\bibnamefont {Jaime}}, \bibinfo {author}
  {\bibfnamefont {N.}~\bibnamefont {Harrison}}, \bibinfo {author}
  {\bibfnamefont {J.~D.}\ \bibnamefont {Thompson}}, \bibinfo {author}
  {\bibfnamefont {V.}~\bibnamefont {Kiryukhin}}, \bibinfo {author}
  {\bibfnamefont {H.~T.}\ \bibnamefont {Yi}}, \bibinfo {author} {\bibfnamefont
  {Y.~S.}\ \bibnamefont {Oh}}, \bibinfo {author} {\bibfnamefont {S.-W.}\
  \bibnamefont {Cheong}}, \bibinfo {author} {\bibfnamefont {C.~D.}\
  \bibnamefont {Batista}}, \ and\ \bibinfo {author} {\bibfnamefont {V.~S.}\
  \bibnamefont {Zapf}},\ }\href@noop {} {\bibfield  {journal} {\bibinfo
  {journal} {Phys. Rev. B}\ }\textbf {\bibinfo {volume} {89}},\ \bibinfo
  {pages} {060404(R)} (\bibinfo {year} {2014})}\BibitemShut {NoStop}%
\bibitem [{\citenamefont {Oh}\ \emph {et~al.}(2014)\citenamefont {Oh},
  \citenamefont {Artyukhin}, \citenamefont {Yang}, \citenamefont {Zapf},
  \citenamefont {Kim}, \citenamefont {Vanderbilt},\ and\ \citenamefont
  {Cheong}}]{Oh2014}%
  \BibitemOpen
  \bibfield  {author} {\bibinfo {author} {\bibfnamefont {Y.~S.}\ \bibnamefont
  {Oh}}, \bibinfo {author} {\bibfnamefont {S.}~\bibnamefont {Artyukhin}},
  \bibinfo {author} {\bibfnamefont {J.~J.}\ \bibnamefont {Yang}}, \bibinfo
  {author} {\bibfnamefont {V.}~\bibnamefont {Zapf}}, \bibinfo {author}
  {\bibfnamefont {J.~W.}\ \bibnamefont {Kim}}, \bibinfo {author} {\bibfnamefont
  {D.}~\bibnamefont {Vanderbilt}}, \ and\ \bibinfo {author} {\bibfnamefont
  {S.-W.}\ \bibnamefont {Cheong}},\ }\href@noop {} {\bibfield  {journal}
  {\bibinfo  {journal} {Nat. Commun.}\ }\textbf {\bibinfo {volume} {5}},\
  \bibinfo {pages} {3201} (\bibinfo {year} {2014})}\BibitemShut {NoStop}%
\bibitem [{\citenamefont {Lee}\ \emph {et~al.}(2013)\citenamefont {Lee},
  \citenamefont {Vecchini}, \citenamefont {Choi}, \citenamefont {Chapon},
  \citenamefont {Bombardi}, \citenamefont {Radaelli},\ and\ \citenamefont
  {Cheong}}]{Lee2013}%
  \BibitemOpen
  \bibfield  {author} {\bibinfo {author} {\bibfnamefont {N.}~\bibnamefont
  {Lee}}, \bibinfo {author} {\bibfnamefont {C.}~\bibnamefont {Vecchini}},
  \bibinfo {author} {\bibfnamefont {Y.~J.}\ \bibnamefont {Choi}}, \bibinfo
  {author} {\bibfnamefont {L.~C.}\ \bibnamefont {Chapon}}, \bibinfo {author}
  {\bibfnamefont {A.}~\bibnamefont {Bombardi}}, \bibinfo {author}
  {\bibfnamefont {P.~G.}\ \bibnamefont {Radaelli}}, \ and\ \bibinfo {author}
  {\bibfnamefont {S.-W.}\ \bibnamefont {Cheong}},\ }\href@noop {} {\bibfield
  {journal} {\bibinfo  {journal} {Phys. Rev. Lett.}\ }\textbf {\bibinfo
  {volume} {110}},\ \bibinfo {pages} {137203} (\bibinfo {year}
  {2013})}\BibitemShut {NoStop}%
\bibitem [{\citenamefont {Aoyama}\ \emph {et~al.}(2014)\citenamefont {Aoyama},
  \citenamefont {Yamauchi}, \citenamefont {Iyama}, \citenamefont {Picozzi},
  \citenamefont {Shimizu},\ and\ \citenamefont {Kimura}}]{Aoyama2014}%
  \BibitemOpen
  \bibfield  {author} {\bibinfo {author} {\bibfnamefont {T.}~\bibnamefont
  {Aoyama}}, \bibinfo {author} {\bibfnamefont {K.}~\bibnamefont {Yamauchi}},
  \bibinfo {author} {\bibfnamefont {A.}~\bibnamefont {Iyama}}, \bibinfo
  {author} {\bibfnamefont {S.}~\bibnamefont {Picozzi}}, \bibinfo {author}
  {\bibfnamefont {K.}~\bibnamefont {Shimizu}}, \ and\ \bibinfo {author}
  {\bibfnamefont {T.}~\bibnamefont {Kimura}},\ }\href@noop {} {\bibfield
  {journal} {\bibinfo  {journal} {Nat. Commun.}\ }\textbf {\bibinfo {volume}
  {5}},\ \bibinfo {pages} {4927} (\bibinfo {year} {2014})}\BibitemShut
  {NoStop}%
\bibitem [{Sup()}]{Supplement}%
  \BibitemOpen
  \href@noop {} {}\bibinfo {note} {See Supplemental Material [url], which
  includes Refs.
  \cite{Detwiler2000,Daou2010,Schmiedeschoff2006,Zapf2010,Wu2010,Blaha2012,Kresse1993,Kresse1994,Blochl1994,Kresse1996a,Kresse1996b,Perdew1996,Perdew1997,Kresse1999,PBE0},
  for the experimental details, poling electric field-dependence of $\Delta
  P(H)$, magnetic field-dependence of ME coefficient, specific heat, and
  details of DFT calculations.}\BibitemShut {Stop}%
\bibitem [{\citenamefont {Chun}\ \emph {et~al.}(2010)\citenamefont {Chun},
  \citenamefont {Chai}, \citenamefont {Oh}, \citenamefont {Jaiswal-Nagar},
  \citenamefont {Haam}, \citenamefont {Kim}, \citenamefont {Lee}, \citenamefont
  {Nam}, \citenamefont {Ko}, \citenamefont {Park}, \citenamefont {Chung},\ and\
  \citenamefont {Kim}}]{Chun2010}%
  \BibitemOpen
  \bibfield  {author} {\bibinfo {author} {\bibfnamefont {S.~H.}\ \bibnamefont
  {Chun}}, \bibinfo {author} {\bibfnamefont {Y.~S.}\ \bibnamefont {Chai}},
  \bibinfo {author} {\bibfnamefont {Y.~S.}\ \bibnamefont {Oh}}, \bibinfo
  {author} {\bibfnamefont {D.}~\bibnamefont {Jaiswal-Nagar}}, \bibinfo {author}
  {\bibfnamefont {S.~Y.}\ \bibnamefont {Haam}}, \bibinfo {author}
  {\bibfnamefont {I.}~\bibnamefont {Kim}}, \bibinfo {author} {\bibfnamefont
  {B.}~\bibnamefont {Lee}}, \bibinfo {author} {\bibfnamefont {D.~H.}\
  \bibnamefont {Nam}}, \bibinfo {author} {\bibfnamefont {K.-T.}\ \bibnamefont
  {Ko}}, \bibinfo {author} {\bibfnamefont {J.~H.}\ \bibnamefont {Park}},
  \bibinfo {author} {\bibfnamefont {J.-H.}\ \bibnamefont {Chung}}, \ and\
  \bibinfo {author} {\bibfnamefont {K.~H.}\ \bibnamefont {Kim}},\ }\href@noop
  {} {\bibfield  {journal} {\bibinfo  {journal} {Phys. Rev. Lett.}\ }\textbf
  {\bibinfo {volume} {104}},\ \bibinfo {pages} {037204} (\bibinfo {year}
  {2010})}\BibitemShut {NoStop}%
\bibitem [{\citenamefont {Kadomtseva}\ \emph {et~al.}(2004)\citenamefont
  {Kadomtseva}, \citenamefont {Zvezdin}, \citenamefont {Popov}, \citenamefont
  {Pyatakov},\ and\ \citenamefont {Vorobe'v}}]{Kadomtseva2004}%
  \BibitemOpen
  \bibfield  {author} {\bibinfo {author} {\bibfnamefont {A.~M.}\ \bibnamefont
  {Kadomtseva}}, \bibinfo {author} {\bibfnamefont {A.~K.}\ \bibnamefont
  {Zvezdin}}, \bibinfo {author} {\bibfnamefont {Y.~F.}\ \bibnamefont {Popov}},
  \bibinfo {author} {\bibfnamefont {A.~P.}\ \bibnamefont {Pyatakov}}, \ and\
  \bibinfo {author} {\bibfnamefont {G.~P.}\ \bibnamefont {Vorobe'v}},\
  }\href@noop {} {\bibfield  {journal} {\bibinfo  {journal} {JETP Letters}\
  }\textbf {\bibinfo {volume} {79}},\ \bibinfo {pages} {571} (\bibinfo {year}
  {2004})}\BibitemShut {NoStop}%
\bibitem [{\citenamefont {Tokunaga}\ \emph {et~al.}(2010)\citenamefont
  {Tokunaga}, \citenamefont {Azuma},\ and\ \citenamefont
  {Shimakawa}}]{Tokunaga2010}%
  \BibitemOpen
  \bibfield  {author} {\bibinfo {author} {\bibfnamefont {M.}~\bibnamefont
  {Tokunaga}}, \bibinfo {author} {\bibfnamefont {M.}~\bibnamefont {Azuma}}, \
  and\ \bibinfo {author} {\bibfnamefont {Y.}~\bibnamefont {Shimakawa}},\
  }\href@noop {} {\bibfield  {journal} {\bibinfo  {journal} {J. Phys. Soc.
  Jpn.}\ }\textbf {\bibinfo {volume} {79}},\ \bibinfo {pages} {064713}
  (\bibinfo {year} {2010})}\BibitemShut {NoStop}%
\bibitem [{\citenamefont {Ivanov}\ \emph {et~al.}(2013)\citenamefont {Ivanov},
  \citenamefont {Mathieu}, \citenamefont {Nordblad}, \citenamefont {Tellgren},
  \citenamefont {Ritter}, \citenamefont {Politova}, \citenamefont {Kaleva},
  \citenamefont {Mosunov}, \citenamefont {Stefanovich},\ and\ \citenamefont
  {Weil}}]{Ivanov2013b}%
  \BibitemOpen
  \bibfield  {author} {\bibinfo {author} {\bibfnamefont {S.~A.}\ \bibnamefont
  {Ivanov}}, \bibinfo {author} {\bibfnamefont {R.}~\bibnamefont {Mathieu}},
  \bibinfo {author} {\bibfnamefont {P.}~\bibnamefont {Nordblad}}, \bibinfo
  {author} {\bibfnamefont {R.}~\bibnamefont {Tellgren}}, \bibinfo {author}
  {\bibfnamefont {C.}~\bibnamefont {Ritter}}, \bibinfo {author} {\bibfnamefont
  {E.}~\bibnamefont {Politova}}, \bibinfo {author} {\bibfnamefont
  {G.}~\bibnamefont {Kaleva}}, \bibinfo {author} {\bibfnamefont
  {A.}~\bibnamefont {Mosunov}}, \bibinfo {author} {\bibfnamefont
  {S.}~\bibnamefont {Stefanovich}}, \ and\ \bibinfo {author} {\bibfnamefont
  {M.}~\bibnamefont {Weil}},\ }\href@noop {} {\bibfield  {journal} {\bibinfo
  {journal} {Chem. Mater.}\ }\textbf {\bibinfo {volume} {25}},\ \bibinfo
  {pages} {935} (\bibinfo {year} {2013})}\BibitemShut {NoStop}%
\bibitem [{\citenamefont {Wu}\ \emph {et~al.}(2010)\citenamefont {Wu},
  \citenamefont {Kan}, \citenamefont {Tian},\ and\ \citenamefont
  {Whangbo}}]{Wu2010}%
  \BibitemOpen
  \bibfield  {author} {\bibinfo {author} {\bibfnamefont {F.}~\bibnamefont
  {Wu}}, \bibinfo {author} {\bibfnamefont {E.}~\bibnamefont {Kan}}, \bibinfo
  {author} {\bibfnamefont {C.}~\bibnamefont {Tian}}, \ and\ \bibinfo {author}
  {\bibfnamefont {M.-H.}\ \bibnamefont {Whangbo}},\ }\href@noop {} {\bibfield
  {journal} {\bibinfo  {journal} {Inorg. Chem.}\ }\textbf {\bibinfo {volume}
  {49}},\ \bibinfo {pages} {7545} (\bibinfo {year} {2010})}\BibitemShut
  {NoStop}%
\bibitem [{\citenamefont {Adamo}\ and\ \citenamefont {Barone}(1999)}]{PBE0}%
  \BibitemOpen
  \bibfield  {author} {\bibinfo {author} {\bibfnamefont {C.}~\bibnamefont
  {Adamo}}\ and\ \bibinfo {author} {\bibfnamefont {V.}~\bibnamefont {Barone}},\
  }\href@noop {} {\bibfield  {journal} {\bibinfo  {journal} {J. Chem. Phys.}\
  }\textbf {\bibinfo {volume} {110}},\ \bibinfo {pages} {6158} (\bibinfo {year}
  {1999})}\BibitemShut {NoStop}%
\bibitem [{\citenamefont {Landau}\ \emph {et~al.}(1984)\citenamefont {Landau},
  \citenamefont {Lifshitz},\ and\ \citenamefont {Pitaevskii}}]{LL8c50}%
  \BibitemOpen
  \bibfield  {author} {\bibinfo {author} {\bibfnamefont {L.~D.}\ \bibnamefont
  {Landau}}, \bibinfo {author} {\bibfnamefont {E.~M.}\ \bibnamefont
  {Lifshitz}}, \ and\ \bibinfo {author} {\bibfnamefont {L.~P.}\ \bibnamefont
  {Pitaevskii}},\ }\href@noop {} {\emph {\bibinfo {title} {Electrodynamics of
  Continuous Media, chapter 50}}},\ \bibinfo {edition} {2nd}\ ed.,\
  Vol.~\bibinfo {volume} {8}\ (\bibinfo  {publisher} {Butterworth-Heinemann},\
  \bibinfo {year} {1984})\BibitemShut {NoStop}%
\bibitem [{\citenamefont {Ivanov}\ \emph {et~al.}(2011)\citenamefont {Ivanov},
  \citenamefont {Nordblad}, \citenamefont {Mathieu}, \citenamefont {Tellgren},
  \citenamefont {Ritter}, \citenamefont {Golubko}, \citenamefont {Politova},\
  and\ \citenamefont {Weil}}]{Ivanov2011}%
  \BibitemOpen
  \bibfield  {author} {\bibinfo {author} {\bibfnamefont {S.~A.}\ \bibnamefont
  {Ivanov}}, \bibinfo {author} {\bibfnamefont {P.}~\bibnamefont {Nordblad}},
  \bibinfo {author} {\bibfnamefont {R.}~\bibnamefont {Mathieu}}, \bibinfo
  {author} {\bibfnamefont {R.}~\bibnamefont {Tellgren}}, \bibinfo {author}
  {\bibfnamefont {C.}~\bibnamefont {Ritter}}, \bibinfo {author} {\bibfnamefont
  {N.~V.}\ \bibnamefont {Golubko}}, \bibinfo {author} {\bibfnamefont {E.~D.}\
  \bibnamefont {Politova}}, \ and\ \bibinfo {author} {\bibfnamefont
  {M.}~\bibnamefont {Weil}},\ }\href@noop {} {\bibfield  {journal} {\bibinfo
  {journal} {Mater. Res. Bull}\ }\textbf {\bibinfo {volume} {46}},\ \bibinfo
  {pages} {1870} (\bibinfo {year} {2011})}\BibitemShut {NoStop}%
\bibitem [{\citenamefont {Hudl}\ \emph {et~al.}(2011)\citenamefont {Hudl},
  \citenamefont {Mathieu}, \citenamefont {Ivanov}, \citenamefont {Weil},
  \citenamefont {Carolus}, \citenamefont {Lottermoser}, \citenamefont {Fiebig},
  \citenamefont {Tokunaga}, \citenamefont {Taguchi}, \citenamefont {Tokura},\
  and\ \citenamefont {Nordblad}}]{Hudl2011}%
  \BibitemOpen
  \bibfield  {author} {\bibinfo {author} {\bibfnamefont {M.}~\bibnamefont
  {Hudl}}, \bibinfo {author} {\bibfnamefont {R.}~\bibnamefont {Mathieu}},
  \bibinfo {author} {\bibfnamefont {S.~A.}\ \bibnamefont {Ivanov}}, \bibinfo
  {author} {\bibfnamefont {M.}~\bibnamefont {Weil}}, \bibinfo {author}
  {\bibfnamefont {V.}~\bibnamefont {Carolus}}, \bibinfo {author} {\bibfnamefont
  {T.}~\bibnamefont {Lottermoser}}, \bibinfo {author} {\bibfnamefont
  {M.}~\bibnamefont {Fiebig}}, \bibinfo {author} {\bibfnamefont
  {Y.}~\bibnamefont {Tokunaga}}, \bibinfo {author} {\bibfnamefont
  {Y.}~\bibnamefont {Taguchi}}, \bibinfo {author} {\bibfnamefont
  {Y.}~\bibnamefont {Tokura}}, \ and\ \bibinfo {author} {\bibfnamefont
  {P.}~\bibnamefont {Nordblad}},\ }\href@noop {} {\bibfield  {journal}
  {\bibinfo  {journal} {Phys. Rev. B}\ }\textbf {\bibinfo {volume} {84}},\
  \bibinfo {pages} {180404(R)} (\bibinfo {year} {2011})}\BibitemShut {NoStop}%
\bibitem [{\citenamefont {Li}\ \emph {et~al.}(2012)\citenamefont {Li},
  \citenamefont {Wang}, \citenamefont {Hsu}, \citenamefont {Lee}, \citenamefont
  {Wu}, \citenamefont {Chou}, \citenamefont {Yang}, \citenamefont {Zhao},
  \citenamefont {Chang}, \citenamefont {Lynn},\ and\ \citenamefont
  {Berger}}]{Li2012}%
  \BibitemOpen
  \bibfield  {author} {\bibinfo {author} {\bibfnamefont {W.~H.}\ \bibnamefont
  {Li}}, \bibinfo {author} {\bibfnamefont {C.~W.}\ \bibnamefont {Wang}},
  \bibinfo {author} {\bibfnamefont {D.}~\bibnamefont {Hsu}}, \bibinfo {author}
  {\bibfnamefont {C.~H.}\ \bibnamefont {Lee}}, \bibinfo {author} {\bibfnamefont
  {C.~M.}\ \bibnamefont {Wu}}, \bibinfo {author} {\bibfnamefont {C.~C.}\
  \bibnamefont {Chou}}, \bibinfo {author} {\bibfnamefont {H.~D.}\ \bibnamefont
  {Yang}}, \bibinfo {author} {\bibfnamefont {Y.}~\bibnamefont {Zhao}}, \bibinfo
  {author} {\bibfnamefont {S.}~\bibnamefont {Chang}}, \bibinfo {author}
  {\bibfnamefont {J.~W.}\ \bibnamefont {Lynn}}, \ and\ \bibinfo {author}
  {\bibfnamefont {H.}~\bibnamefont {Berger}},\ }\href@noop {} {\bibfield
  {journal} {\bibinfo  {journal} {Phys. Rev. B}\ }\textbf {\bibinfo {volume}
  {85}},\ \bibinfo {pages} {094431} (\bibinfo {year} {2012})}\BibitemShut
  {NoStop}%
\bibitem [{\citenamefont {Detwiler}\ \emph {et~al.}(2000)\citenamefont
  {Detwiler}, \citenamefont {Schmiedeshoff}, \citenamefont {Harrison},
  \citenamefont {Lacerda}, \citenamefont {Cooley},\ and\ \citenamefont
  {Smith}}]{Detwiler2000}%
  \BibitemOpen
  \bibfield  {author} {\bibinfo {author} {\bibfnamefont {J.~A.}\ \bibnamefont
  {Detwiler}}, \bibinfo {author} {\bibfnamefont {G.~M.}\ \bibnamefont
  {Schmiedeshoff}}, \bibinfo {author} {\bibfnamefont {N.}~\bibnamefont
  {Harrison}}, \bibinfo {author} {\bibfnamefont {A.~H.}\ \bibnamefont
  {Lacerda}}, \bibinfo {author} {\bibfnamefont {J.~C.}\ \bibnamefont {Cooley}},
  \ and\ \bibinfo {author} {\bibfnamefont {J.~L.}\ \bibnamefont {Smith}},\
  }\href {\doibase 10.1103/PhysRevB.61.402} {\bibfield  {journal} {\bibinfo
  {journal} {Phys. Rev. B}\ }\textbf {\bibinfo {volume} {61}},\ \bibinfo
  {pages} {402} (\bibinfo {year} {2000})}\BibitemShut {NoStop}%
\bibitem [{\citenamefont {Daou}\ \emph {et~al.}(2010)\citenamefont {Daou},
  \citenamefont {Weickert}, \citenamefont {Nicklas}, \citenamefont {Steglich},
  \citenamefont {Haase},\ and\ \citenamefont {Doerr}}]{Daou2010}%
  \BibitemOpen
  \bibfield  {author} {\bibinfo {author} {\bibfnamefont {R.}~\bibnamefont
  {Daou}}, \bibinfo {author} {\bibfnamefont {F.}~\bibnamefont {Weickert}},
  \bibinfo {author} {\bibfnamefont {M.}~\bibnamefont {Nicklas}}, \bibinfo
  {author} {\bibfnamefont {F.}~\bibnamefont {Steglich}}, \bibinfo {author}
  {\bibfnamefont {A.}~\bibnamefont {Haase}}, \ and\ \bibinfo {author}
  {\bibfnamefont {M.}~\bibnamefont {Doerr}},\ }\href@noop {} {\bibfield
  {journal} {\bibinfo  {journal} {Rev. Sci. Inst.}\ }\textbf {\bibinfo {volume}
  {81}},\ \bibinfo {pages} {033909} (\bibinfo {year} {2010})}\BibitemShut
  {NoStop}%
\bibitem [{\citenamefont {Schmiedeshoff}\ \emph {et~al.}(2006)\citenamefont
  {Schmiedeshoff}, \citenamefont {Lounsbury}, \citenamefont {Luna},
  \citenamefont {Tracy}, \citenamefont {Schramm}, \citenamefont {Tozer},
  \citenamefont {Correa}, \citenamefont {Hannahs}, \citenamefont {Murphy},
  \citenamefont {Palm}, \citenamefont {Lacerda}, \citenamefont {Bud’ko},
  \citenamefont {Canfield}, \citenamefont {Smith}, \citenamefont {Lashley},\
  and\ \citenamefont {Cooley}}]{Schmiedeschoff2006}%
  \BibitemOpen
  \bibfield  {author} {\bibinfo {author} {\bibfnamefont {G.~M.}\ \bibnamefont
  {Schmiedeshoff}}, \bibinfo {author} {\bibfnamefont {A.~W.}\ \bibnamefont
  {Lounsbury}}, \bibinfo {author} {\bibfnamefont {D.~J.}\ \bibnamefont {Luna}},
  \bibinfo {author} {\bibfnamefont {S.~J.}\ \bibnamefont {Tracy}}, \bibinfo
  {author} {\bibfnamefont {A.~J.}\ \bibnamefont {Schramm}}, \bibinfo {author}
  {\bibfnamefont {S.~W.}\ \bibnamefont {Tozer}}, \bibinfo {author}
  {\bibfnamefont {V.~F.}\ \bibnamefont {Correa}}, \bibinfo {author}
  {\bibfnamefont {S.~T.}\ \bibnamefont {Hannahs}}, \bibinfo {author}
  {\bibfnamefont {T.~P.}\ \bibnamefont {Murphy}}, \bibinfo {author}
  {\bibfnamefont {E.~C.}\ \bibnamefont {Palm}}, \bibinfo {author}
  {\bibfnamefont {A.~H.}\ \bibnamefont {Lacerda}}, \bibinfo {author}
  {\bibfnamefont {S.~L.}\ \bibnamefont {Bud’ko}}, \bibinfo {author}
  {\bibfnamefont {P.~C.}\ \bibnamefont {Canfield}}, \bibinfo {author}
  {\bibfnamefont {J.~L.}\ \bibnamefont {Smith}}, \bibinfo {author}
  {\bibfnamefont {J.~C.}\ \bibnamefont {Lashley}}, \ and\ \bibinfo {author}
  {\bibfnamefont {J.~C.}\ \bibnamefont {Cooley}},\ }\href@noop {} {\bibfield
  {journal} {\bibinfo  {journal} {Rev. Sci. Instrum.}\ }\textbf {\bibinfo
  {volume} {77}},\ \bibinfo {pages} {123907} (\bibinfo {year}
  {2006})}\BibitemShut {NoStop}%
\bibitem [{\citenamefont {Zapf}\ \emph {et~al.}(2010)\citenamefont {Zapf},
  \citenamefont {Kenzelmann}, \citenamefont {Wolff-Fabris}, \citenamefont
  {Balakirev},\ and\ \citenamefont {Chen}}]{Zapf2010}%
  \BibitemOpen
  \bibfield  {author} {\bibinfo {author} {\bibfnamefont {V.~S.}\ \bibnamefont
  {Zapf}}, \bibinfo {author} {\bibfnamefont {M.}~\bibnamefont {Kenzelmann}},
  \bibinfo {author} {\bibfnamefont {F.}~\bibnamefont {Wolff-Fabris}}, \bibinfo
  {author} {\bibfnamefont {F.}~\bibnamefont {Balakirev}}, \ and\ \bibinfo
  {author} {\bibfnamefont {Y.}~\bibnamefont {Chen}},\ }\href@noop {} {\bibfield
   {journal} {\bibinfo  {journal} {Phys. Rev. B}\ }\textbf {\bibinfo {volume}
  {82}},\ \bibinfo {pages} {060402} (\bibinfo {year} {2010})}\BibitemShut
  {NoStop}%
\bibitem [{\citenamefont {Rocquefelte}\ \emph {et~al.}(2012)\citenamefont
  {Rocquefelte}, \citenamefont {Schwarz},\ and\ \citenamefont
  {Blaha}}]{Blaha2012}%
  \BibitemOpen
  \bibfield  {author} {\bibinfo {author} {\bibfnamefont {X.}~\bibnamefont
  {Rocquefelte}}, \bibinfo {author} {\bibfnamefont {K.}~\bibnamefont
  {Schwarz}}, \ and\ \bibinfo {author} {\bibfnamefont {P.}~\bibnamefont
  {Blaha}},\ }\href@noop {} {\bibfield  {journal} {\bibinfo  {journal} {Sci.
  Rep.}\ }\textbf {\bibinfo {volume} {2}},\ \bibinfo {pages} {759} (\bibinfo
  {year} {2012})}\BibitemShut {NoStop}%
\bibitem [{\citenamefont {Kresse}\ and\ \citenamefont
  {Hafner}(1993)}]{Kresse1993}%
  \BibitemOpen
  \bibfield  {author} {\bibinfo {author} {\bibfnamefont {G.}~\bibnamefont
  {Kresse}}\ and\ \bibinfo {author} {\bibfnamefont {J.}~\bibnamefont
  {Hafner}},\ }\href@noop {} {\bibfield  {journal} {\bibinfo  {journal} {Phys.
  Rev. B}\ }\textbf {\bibinfo {volume} {47}},\ \bibinfo {pages} {558} (\bibinfo
  {year} {1993})}\BibitemShut {NoStop}%
\bibitem [{\citenamefont {Kresse}\ and\ \citenamefont
  {Hafner}(1994)}]{Kresse1994}%
  \BibitemOpen
  \bibfield  {author} {\bibinfo {author} {\bibfnamefont {G.}~\bibnamefont
  {Kresse}}\ and\ \bibinfo {author} {\bibfnamefont {J.}~\bibnamefont
  {Hafner}},\ }\href@noop {} {\bibfield  {journal} {\bibinfo  {journal} {Phys.
  Rev. B}\ }\textbf {\bibinfo {volume} {49}},\ \bibinfo {pages} {14251}
  (\bibinfo {year} {1994})}\BibitemShut {NoStop}%
\bibitem [{\citenamefont {Blochl}(1994)}]{Blochl1994}%
  \BibitemOpen
  \bibfield  {author} {\bibinfo {author} {\bibfnamefont {P.~E.}\ \bibnamefont
  {Blochl}},\ }\href@noop {} {\bibfield  {journal} {\bibinfo  {journal} {Phys.
  Rev. B}\ }\textbf {\bibinfo {volume} {50}},\ \bibinfo {pages} {17953}
  (\bibinfo {year} {1994})}\BibitemShut {NoStop}%
\bibitem [{\citenamefont {Kresse}\ and\ \citenamefont
  {Furthmller}(1996)}]{Kresse1996a}%
  \BibitemOpen
  \bibfield  {author} {\bibinfo {author} {\bibfnamefont {G.}~\bibnamefont
  {Kresse}}\ and\ \bibinfo {author} {\bibfnamefont {J.}~\bibnamefont
  {Furthmller}},\ }\href@noop {} {\bibfield  {journal} {\bibinfo  {journal}
  {Comput. Mat. Sci.}\ }\textbf {\bibinfo {volume} {6}},\ \bibinfo {pages} {15}
  (\bibinfo {year} {1996})}\BibitemShut {NoStop}%
\bibitem [{\citenamefont {Kresse}\ and\ \citenamefont
  {Furthmuller}(1996)}]{Kresse1996b}%
  \BibitemOpen
  \bibfield  {author} {\bibinfo {author} {\bibfnamefont {G.}~\bibnamefont
  {Kresse}}\ and\ \bibinfo {author} {\bibfnamefont {J.}~\bibnamefont
  {Furthmuller}},\ }\href@noop {} {\bibfield  {journal} {\bibinfo  {journal}
  {Phys. Rev. B}\ }\textbf {\bibinfo {volume} {54}},\ \bibinfo {pages} {11169}
  (\bibinfo {year} {1996})}\BibitemShut {NoStop}%
\bibitem [{\citenamefont {Perdew}\ \emph {et~al.}(1996)\citenamefont {Perdew},
  \citenamefont {Burke},\ and\ \citenamefont {Ernzerhof}}]{Perdew1996}%
  \BibitemOpen
  \bibfield  {author} {\bibinfo {author} {\bibfnamefont {J.~P.}\ \bibnamefont
  {Perdew}}, \bibinfo {author} {\bibfnamefont {K.}~\bibnamefont {Burke}}, \
  and\ \bibinfo {author} {\bibfnamefont {M.}~\bibnamefont {Ernzerhof}},\
  }\href@noop {} {\bibfield  {journal} {\bibinfo  {journal} {Phys. Rev. Lett.}\
  }\textbf {\bibinfo {volume} {77}},\ \bibinfo {pages} {3865} (\bibinfo {year}
  {1996})}\BibitemShut {NoStop}%
\bibitem [{\citenamefont {Perdew}\ \emph {et~al.}(1997)\citenamefont {Perdew},
  \citenamefont {Burke},\ and\ \citenamefont {Ernzerhof}}]{Perdew1997}%
  \BibitemOpen
  \bibfield  {author} {\bibinfo {author} {\bibfnamefont {J.~P.}\ \bibnamefont
  {Perdew}}, \bibinfo {author} {\bibfnamefont {K.}~\bibnamefont {Burke}}, \
  and\ \bibinfo {author} {\bibfnamefont {M.}~\bibnamefont {Ernzerhof}},\
  }\href@noop {} {\bibfield  {journal} {\bibinfo  {journal} {Phys. Rev. Lett.}\
  }\textbf {\bibinfo {volume} {78}},\ \bibinfo {pages} {1396} (\bibinfo {year}
  {1997})}\BibitemShut {NoStop}%
\bibitem [{\citenamefont {Kresse}\ and\ \citenamefont
  {Joubert}(1999)}]{Kresse1999}%
  \BibitemOpen
  \bibfield  {author} {\bibinfo {author} {\bibfnamefont {G.}~\bibnamefont
  {Kresse}}\ and\ \bibinfo {author} {\bibfnamefont {D.}~\bibnamefont
  {Joubert}},\ }\href@noop {} {\bibfield  {journal} {\bibinfo  {journal} {Phys.
  Rev. B}\ }\textbf {\bibinfo {volume} {59}},\ \bibinfo {pages} {1758}
  (\bibinfo {year} {1999})}\BibitemShut {NoStop}%
\end{thebibliography}%

\end{document}